\documentclass[final]{OAGM}
\OAGMarXiv{1404.3538}
\pdfoutput=1
\usepackage{graphicx}
\usepackage{multicol}
\usepackage{caption}
\usepackage{floatflt}
\usepackage{subcaption}

\title{ILATO Project:  Fusion of Optical Surface Models and Volumetric CT Data}

\author{Andreas Beyer, Hubert Mara, Susanne Kr\"omker\\
Interdisciplinary Center for Scientific Computing (IWR)\\
Heidelberg University, Germany}

\begin{document}
\maketitle

\begin{abstract}
Project ILATO focuses on \textit{ \underline{I}mproving \underline{L}imited \underline{A}ngle computed \underline{T}omography by \underline{O}ptical data integration} in order to enhance image quality and shorten acquisition times in X-ray based industrial quality inspection. Limited angle computed tomography is indicated whenever specimen dimensions exceed cone beam limits or the object is impenetrable from certain angles. Thus, acquiring only a subset of a full circle CT scan poses problems for reconstruction algorithms due to incomplete data which introduces blurred edges and other artifacts. To support volumetric data reconstruction algorithm a surface mesh of the object obtained via structured light optical scan acts as a mask defining boundaries of the reconstructed image. The registration of optically acquired surfaces with data acquired from computed tomography is our current challenge. This article presents our setup, the methods applied and discusses the problems arising from registration of data sets created with considerably different imaging techniques.
\end{abstract}

\section{Introduction}\label{introduction}
In the field of industrial quality inspection tactile measurements are more and more superseded by non-destructive measurement methods. Approaches like structured light 3D scanners \cite{Sablatnig1992} or laser based distance measurements generate highly accurate representations of the test object\rq{}s surface. During this process an irregular triangular mesh of the object\rq{}s surface with a spacial resolution down to $\sim$10$\mu m$ is created. The time needed to acquire this information varies between seconds and minutes but does not provide interior structures. As an alternative method to discover interior damages or indication of wear out computed tomography (CT) scans create volumetric data as a digital representation of internal structures with a resolution of $\sim$75$\mu m$. The time needed to complete a CT scan is significantly higher than its optical counterpart and can take hours or days depending on size and density of specimen. All those models can be matched against existing digital descriptions used during production such as CAD files, if available. The ILATO project -- a joint project with \textit{Empa - Swiss Federal Laboratories for Materials Science and Technology} -- investigates opportunities to speed up acquisition times in limited angle computed tomography, to optimize CT trajectories and to improve quality of volumetric representation.

The paper is structured as follows: The next section presents the image acquisition techniques Computed Tomography and Optical Imaging. Section~\ref{problem} describes challenges from combining output of different image acquisition techniques. The final section summarizes the current state of the project and gives an outlook to future work.

\section{Combining Different Imaging Techniques}\label{goal}
X-ray CT offers a non-destructive method for the imaging, inspection and metrology of interior and exterior structures of a specimen. For specimens deviating substantially from circular symmetry modified scan procedures under the name \lq\lq{}Limited Angle CT\rq\rq{} have been suggested, where only a limited number of projections is recorded. It is desired to obtain measurements along a $180^\circ$ circular trajectory, if this is not feasible due to highly anisotropic shaped objects or impenetrable regions, Limited Angle CT is indicated only collecting a subset of all projections e.g.~on arcs of less than $180^\circ$. Due to the limited information, the accuracy of the measurement varies significantly over the specimen\cite{Schuetz2013}. The optical scan is not limited in acquisition angles but in the contrary suffers from shadowing effects. This work focuses on the registration process with volumetric data sets in order to generate masks as complementary data from optical surface models and estimate weights for reconstruction methods explained in section~\ref{analyticalReconstruction} to mitigates the artifacts introduceb by Limited Angle CT.

A fundamental requirements of these applications is the estimation of scene depth information or representation of recorded structures, preferably in real time, with infinite resolution, contrast and without noise. It is not possible to achieve all of this, and the hardware setup of our project is fixed. The following sections introduce basic principles of techniques occuring in our setup. CT hardware and related reconstruction is in the project partner\rq{}s responsibility, they pursue a filtered backprojection approach. Our optical scanning system combines both approaches presented in section~\ref{opticalScan}.

\subsection{Computed Tomography Reconstruction Methods}\label{computedTomography}
Standard CT workflow is to acquire a series of attenuation measurements of the specimen from various angles and compute a volumetric model. The object under investigation is placed on a rotating table between X-ray source and detector screen. From a finite number of projections of an object, mostly on circular trajectory, sinograms representing attenuation values of specimen are acquired. 3D reconstruction from a series of 2D projections is accomplished via inverse radon transformation \footnote{Johann Radon. Über die Bestimmung von Funktionen durch ihre Integralwerte längs gewisser Mannigfaltigkeiten. Akad. Wiss., 69:262–277, 1917}. To perform the reconstruction task iterative or analytical algorithms are known as described below.

\subsubsection*{Iterative Reconstruction}
is a fundamentally different approach for obtaining the volumetric representation from sinograms \cite{Hutton2002}. The problem is discretized and an approximate solution is computed by iterative methods. In the so-called Algebraic Reconstruction Technique (ART), projections are discretized yielding a huge linear system of equations. Statistical reconstruction methods otherwise take into account the random nature of the measurements; they are based on the minimization of the distance between the measured data and the estimations given by a statistical model. Iterative reconstruction is computationally expensive compared to analytical approach and tends to increase resolution. The drawback is that each iteration amplifies noise. 

\subsubsection*{Analytical Reconstruction}\label{analyticalReconstruction}
is commonly performed as filtered backprojection which consists of two steps: At first, sinograms are run back through the image to obtain a rough approximation to the original. Since this causes some blurring in other parts of the reconstructed image the second step is to apply a deconvolution filter or ramp filter per projection to mitigate the blurring effects. Systems differ in choice of applied filter and order of steps but in general follow this idea. Analytical methods have the advantage to be fast and deliver good quality results under standard scanning conditions\cite{Montes2007}. To enhance filtered backprojection approach a mask defining boundaries of the object can be included in reconstruction algorithm. This is called weighted filtered backprojection. 

\subsection{Optical Surface Scan}\label{opticalScan}
Surface scanning refers to optical systems that measure objects through visible light and generate dense 3D polygonal meshes. The resulting data are restricted to the outer shell of an object, which distinguishes surface scanning from volume scanning. A major distinction is made with respect to the underlying principle of computing $xyz$ coordinates, the most common being time-of-flight and triangulation. 3D imaging systems can use passive image capture or active light projection. Optical scans in our project are performed with a  \textit{Breuckmann SmartScan 3D-HE} system combining stereoscopic acquisition and fringe light projection. Our partners at Empa are using a \textit{Konica Minolta VI-900} employing a laser line only with structured light acquisition. The resulting 3D models of these optical scanners are processed with GigaMesh\cite{Mara2010} software framework incorporating algorithms of ILATO project as library.

\subsubsection*{Stereoscopic Acquisition}
uses typically two cameras with same focal length which are mounted parallel to each other. Both cameras view the same real-world point in a different location on the 2D images acquired. The projections of a real world point in the left and right camera image have a distance which is known as disparity. This can be used to calculate depth information, which is the distance between the real-world point and the stereo vision system \cite{Koch1995}. 

\subsubsection*{Structured Light Scanners}
use fringe or Moir\'{e} projection and/or phase shift technology. Moir\'{e} patterns are a series of non-random linear projections onto the surface of the object. Multiple captures of the same pattern, slightly shifted, improve the measurement accuracy but increase acquisition times \cite{Bathow2010}. By projecting a regular pattern onto an object and recording the resulting image depth information is obtained and a 3D model can be constructed as point cloud. Further information about the continuity of the surface is represented by connecting points and forming triangles. 

\section{Registration of Dual Sensor Surface Models }\label{problem}
Optically acquired data need to be registered with data from CT to support reconstruction algorithm by applying a mask or weights to filtered backprojection. Attempts for accurate and fast automatic registration are discussed in the following.

\subsection{Challenges}
We have to deal with imbalanced data sets since optical scans provide a fine grain surface structure but CT scans offer a rather coarse grain mesh including interior structures. The optical surface model represents a clipping of the surface extracted from CT model. In the presented example shown in figure \ref{convexhull} the optical model has 5 million vertices to represent a total surface area of $1042~cm^2$, while the volumetric model of the same object represents all interior and exterior structures in 0.5 million vertices covering a total area of $1507~cm^2$ after surface triangulation. Most approaches implicitly consider the meshes to be produced by the very same imaging technique which implies the assumption to work with similar resolution, similar grid regularities and a common set of features. The same implicit assumption applies when operating on synthetic meshes or having full control over the process of transferring a volumetric model to a surface mesh. The resolution numbers given in the introduction certainly only apply for our setup but they indicate that for example distance measurements between identical pairs of points in both mesh representations suddenly yield different values. 

\subsection{Pre-alignment by Global Features}\label{solution}
As initial step and to approximate orientation and placement of optical surface models and volumetric data, per mesh analysis based on surface structure is performed. All efforts to compute a proper alignment of both data sets considering all vertices suffer from imbalanced complexity of models. Optical data represents surface structures in great detail but lacks information about the interior of the object. Computing the convex hull and minimal bounding box have proven to be very reliable for finding a relation between both representations solely on global features. 
\vspace{-10pt}
\subsubsection*{Minimal Enclosing Ellipsoid} 
\begin{figure}[h]
\vspace{-10pt}
\centering
\begin{subfigure}[b]{0.49\textwidth}
\includegraphics[width=0.7\textwidth]{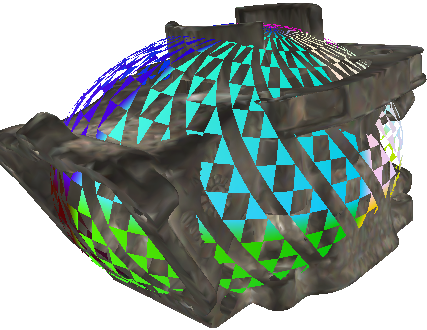}

\caption{}\label{convexhull}
\end{subfigure}
   ~
     \begin{subfigure}[b]{0.49\textwidth}
\includegraphics[width=0.7\textwidth]{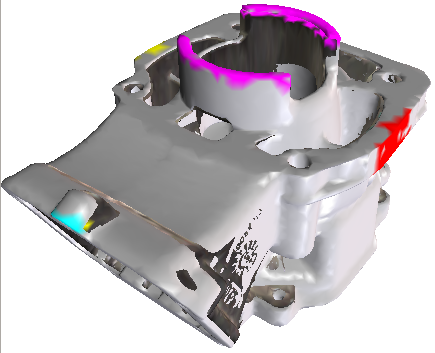}
\caption{}\label{prealignment}
\end{subfigure}
\vspace{-20pt}
\caption{Test object with (a) MEE and (b) preliminary alignment}\label{prealign}
\vspace{-20pt}
\end{figure}

(MEE) is the smallest ellipsoid containing the respective point set analogous to the convex hull \cite{Barber1996}. In 3D case we represent this as an ellipsoid  around the respective object (figure \ref{convexhull} with downscaled MEE to fit image) to estimate dimensions based on surface data. In contrast to this computing the center of gravity or performing Principal Component Analysis (PCA) also includes internal data points only represented in CT mesh and therefore does not provide reliable information to align volumetric scans with surface scans.

\subsubsection*{Minimal Bounding Box} (MBB) is computed to calculate cut plane volume of mesh and the six planes of MBB \cite{Har-Peled2001}. In combination with MEE Analysis, this provides reliable orientation information by comparing MEE center and MBB center. It works as a rough estimation for non rotation symmetric objects as encountered in many real world scenarios. Figure \ref{prealignment} represents the result of approximated alignment based on MBB and MEE. The algorithm was originally intended to calculate the exact diameter of a point set but also can estimate a minimal volume bounding box. As this approach considers only a subset of all points and orders them in a fair-split tree the accuracy of resulting MBB depends on the size of the subset. In our case a fast computed rough estimation of MBB returns acceptable results.

\subsubsection*{Cut Planes}\label{cutPlanes} 
\begin{figure}[h]
\centering
\begin{subfigure}[b]{0.35\textwidth}
\vspace{-20pt}
\includegraphics[width=0.9\textwidth]{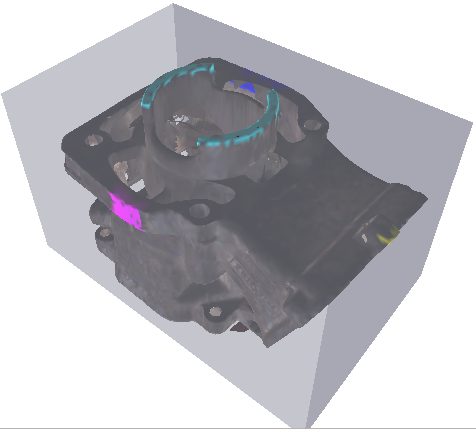}
\caption{}\label{fig:mbb}
\end{subfigure}%
       \qquad~
       \begin{subfigure}[b]{0.35\textwidth}
\includegraphics[width=0.9\textwidth]{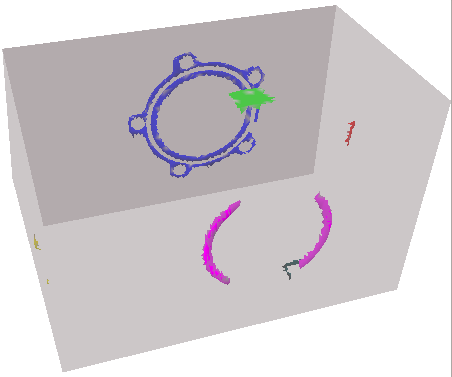}
\caption{}\label{fig:cutsWithMBB}
\end{subfigure} 
\vspace{-10pt}
     \caption{(a) MBB and (b) intersections with MBB}\label{fig:mbbCuts}
\vspace{-15pt}
\end{figure}

from intersections of mesh and all six sides of 1\% shrunk MBB as shown in figure \ref{fig:cutsWithMBB} are extracted. After calculating the surface of all six resulting meshes, the largest of those surfaces is used as input for the registration algorithm. One could also apply the registration process to complete meshes, but all interior structures of CT mesh are not represented in the optical surface mesh and registration complexity significantly rises with the number of vertices. A valid transformation for one part of a rigid body holds for the complete body and this shortens approach to shorten the lengthy registration process.

\subsection{Final Alignment by Local Features}\label{registration}
\begin{figure}[h]
   \vspace{-10pt}
   \centering
        \begin{subfigure}[b]{0.3\textwidth}
                \includegraphics[width=0.8\textwidth]{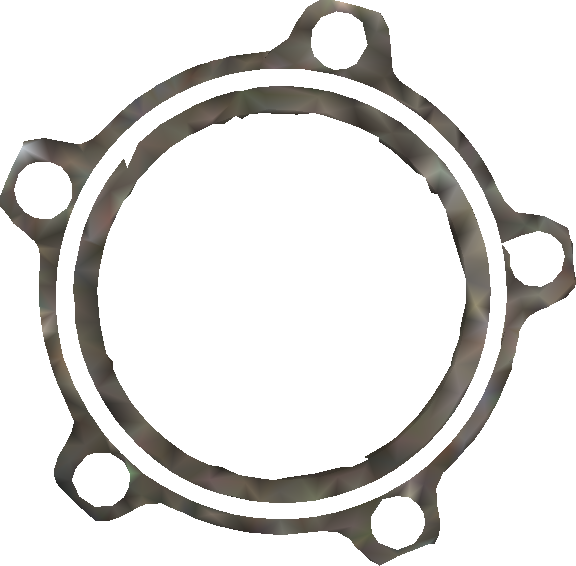}
                \caption{}
                \label{fig:opticalFull}
        \end{subfigure}%
~
        \begin{subfigure}[b]{0.3\textwidth}
                \includegraphics[width=0.8\textwidth]{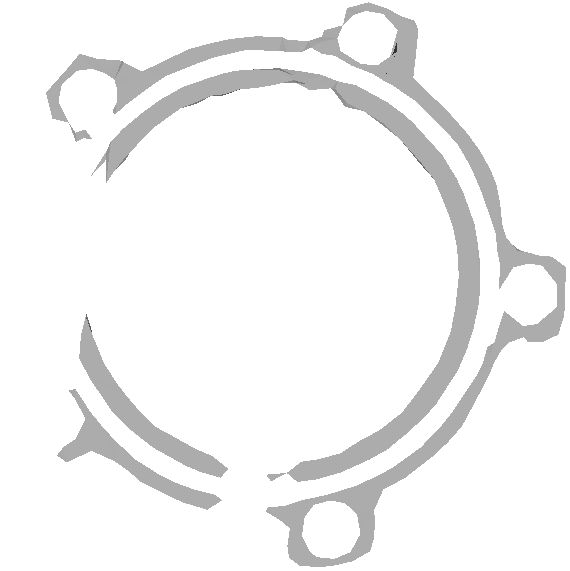}
                \caption{}
                \label{fig:ctPartial}
        \end{subfigure}
~
        \begin{subfigure}[b]{0.3\textwidth}
                \includegraphics[width=0.8\textwidth]{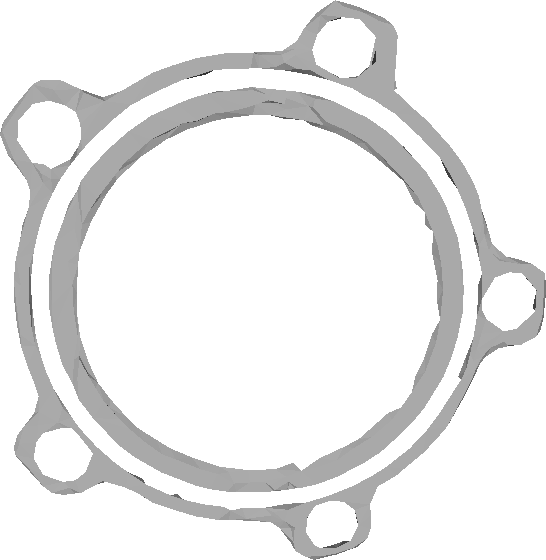}
                \caption{}
                \label{fig:ctFull}
        \end{subfigure}
\vspace{-10pt}
        \caption{(a) full ring from optical scan; (b) partial and (c) full ring from CT}\label{fig:input}
\vspace{-20pt}
\end{figure}

The registration algorithm is applied to the largest intersection of each mesh with its shrunk MBB. Figure \ref{fig:input} shows typical input data for this operation. As seen in figures \ref{fig:opticalFull} and \ref {fig:ctFull} cuts with tight fitting minimal bounding box for our test object returns the bottom plane which is a ring shaped mounting with five screw pits. The MBB is an approximation and therefore results might vary from execution to execution, figure \ref{fig:ctPartial} represents a cut with a not perfectly tight fitting MBB which returns a partial ring structure. In real world scenarios we often encounter this kind of imperfect data.

\subsection{Iterative Closest Point}
The ICP algorithm is generally used to minimize the root mean square (RMS) distance between two point clouds. However, it might converge in an erroneous local minimum, especially for noisy data. To ensure that the solution found by the ICP algorithm is a global minimum, a simulated annealing algorithm can be employed  \cite{Penney2001}. Our project utilizes an ICP algorithm from ITK \cite{Yoo2002} framework with a Levenberg-Marquardt solver to register the meshes described in \ref{cutPlanes}:
\begin{figure}[h]
     \vspace{-10pt}
     \centering
      \begin{subfigure}[b]{0.3\textwidth}
\includegraphics[width=0.8\textwidth]{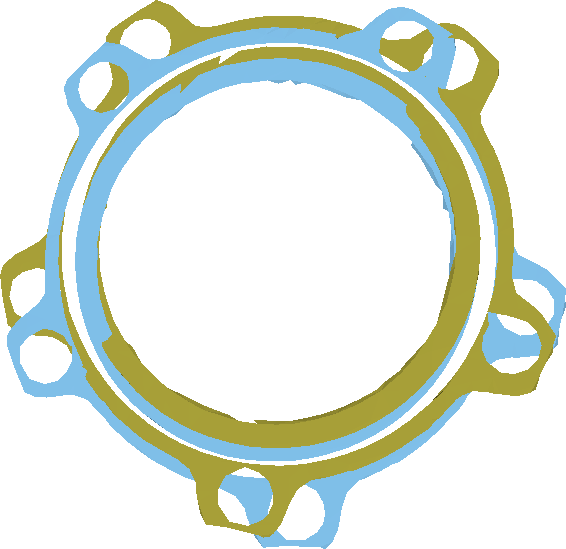}
\caption{}\label{icpFullFull}
        \end{subfigure}
~
  \begin{subfigure}[b]{0.3\textwidth}
\includegraphics[width=0.8\textwidth]{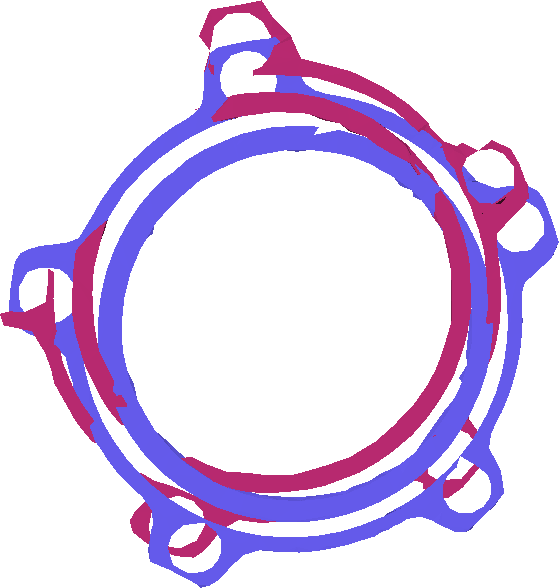}
\caption{}\label{icpFullPart}
        \end{subfigure}
\vspace{-10pt}
 \caption{ICP registration of (a) fig.~\ref{fig:opticalFull} with fig.~\ref{fig:ctFull}, (b) fig.~\ref{fig:opticalFull} with fig.~\ref{fig:ctPartial}}
\label{icpRegistration}
\vspace{-10pt}
\end{figure}
Especially in the case of symmetric shapes ICP is easily trapped in local minima not representing the desired registration as shown in figure \ref{icpFullFull}. Also aligning partial ring structures with full rings often returns faulty matches as shown in figure \ref{icpFullPart} where neither the ring centers nor single screw pits match. Structures as shown in figure \ref{fig:input} are common in machinery parts and therefore subject to quality inspection. ICP registration runs were performed on pre-aligned data as shown in figure \ref{prealign} and on arbitrarily placed input meshes. Runtime is significantly shortened by pre-aligning data but still worse -- also in terms of root mean square error -- compared to our RANSAC approach.

\subsection{RANdom SAmple and Consensus}
In general, RANSAC schemes  generate a random hypothesis to solve a given problem. This hypothesis needs to be evaluated and a quality measure is computed. If the quality measure exceeds a certain threshold the one hypothesis with the best evaluation is returned as solution for the given problem. If not the next hypothesis is generated and evaluated.

Our setups follows an approach as described in \cite{Winkelbach2008} to find a suitable transformation from an optical data mesh to a CT data mesh. In one RANSAC iteration a random pair of points is chosen from the first mesh $A$. Characteristics of this pair such as the distance of points, the inclination of normal vectors and their rotation angle are computed. Now, the corresponding pair and characteristics are stored in a table of mesh $A$ and the same characteristics are searched in the table of mesh $B$. In the next RANSAC iteration the pair is chosen from mesh $B$, characteristics are computed, stored in the table of mesh $B$ and searched in the table of mesh $A$. Once a query returns  successful, the respective point pairs sharing these characteristics are considered a hypothesis. Verification is done by computing a transformation from the pair of $A$ to the pair of $B$ and applying this transformation to various random points from mesh $A$. The more transformed random points have small distances to mesh $B$, the higher the confidence in the current hypothesis is. If the confidence value exceeds a threshold one can assume that the current hypothesis represents a valid registration of both meshes.

\begin{figure}[h]
       \vspace{-10pt}
   \centering
      \begin{subfigure}[b]{0.3\textwidth}
\includegraphics[width=0.8\textwidth]{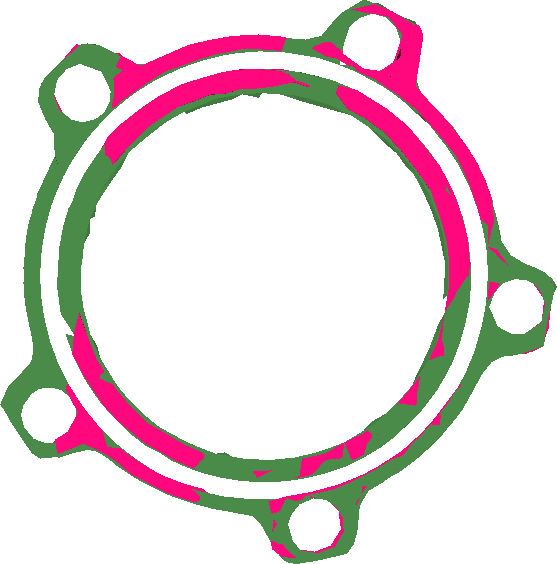}
\caption{}
        \end{subfigure}
~
  \begin{subfigure}[b]{0.3\textwidth}
\includegraphics[width=0.8\textwidth]{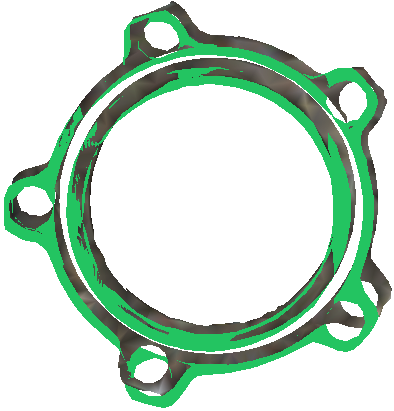}
\caption{}
        \end{subfigure}
 \caption{RANSAC registration of (a) fig.~\ref{fig:opticalFull} with fig.~\ref{fig:ctPartial}, (b) fig.~\ref{fig:opticalFull} with fig.~\ref{fig:ctFull}}
\vspace{-10pt}
\label{ransacRegistration}
\end{figure}
We favor this approach over ICP since it does not suffer from defects as shown in figure \ref{icpRegistration}.  Root mean square deviation (RMSD) minimization tends to shift the partial ring\rq{}s center of gravity towards the full ring\rq{}s center in figure \ref{icpFullPart} which is not desired and can be observed in many similar scenarios.  In addition it is possible to include previous knowledge in RANSAC hypothesis generation and verification.

\subsubsection*{Assessment of registration quality} is usually done by computing the RMSD between transformed mesh and destination mesh. But to measure how far two surfaces in a metric space are from each other Hausdorff Distance \footnote{Felix Hausdorff. Mengenlehre. de Gruyter Berlin, 1927} is more suited. It measures from first to second mesh and vice versa the distance from each point of one mesh to the nearest neighbor in the other mesh and returns the longest distance. The pointwise distance function can be a Manhattan distance also needed in generating ta kd-tree.
To improve point to point distance queries as performed by RANSAC hypothesis verification and Hausdorff measurement, the vertices (per mesh) need to be organized in a space partitioning data structure. An efficient way of implementing searches with a multidimensional search key are kd-trees \cite{Bentley1975} which are a special case of binary space partitioning trees. Storing all vertices of one mesh in such a structure enables nearest neighbor queries in $O(log N)$ time in average case.\vspace{-3pt}

\section{Conclusions and Outlook}\label{conclusions}\vspace{-3pt}
Registering two objects with completely represented surface turns out to be quite a challenge. One model holds significantly more data while the other model provides additional internal structures. Both models come from fundamentally different imaging techniques with different resolution. One holds color information while the other one has significantly varying measurement accuracy over the specimen. ICP suffers from mistakenly compensating interior CT data by shifting the model out of place. RANSAC generates not enough or faulty hypotheses due to mismatch in resolution. We will investigate other metrics to characterize point pairs and incorporate knowledge about accuracy variations in CT model to improve our registration. Once a successful registration is performed during the process of data fusion duplicated vertices will be unified and color information from optical scans are combined with volumetric model.\vspace{-10pt}

\subsubsection*{Acknowledgments}
This project  is financed by the SNF and DFG within the DACH-Framework (Grant-No: BO 864/17-1). The cooperation is carried out with \textit{Empa - Swiss Federal Laboratories for Materials Science and Technology} in D\"ubendorf. We thank our colleagues Alexander Flisch, Philipp Sch\"utz, and Liu Yu from \textit{Empa} for fruitful discussions, CT data acquisition and provision of customized sample specimens. We want to thank the \textit{Heidelberg Graduate School of Mathematical and Computational Methods for the Sciences } for providing an optical scanning system (Breuckmann SmartScan 3D-HE).

\bibliography{beyer_ilato}
\end{document}